\documentclass[prb,aps,twocolumn,amsmath,amsfonts]{revtex4}

\usepackage{graphicx}

\def\revision{}

\begin{document}

\title{Vison excitations in near-critical quantum dimer models}

\author {Gr\'egory Str\"ubi}
\affiliation{Department of Physics, University of Basel,
CH-4056 Basel, Switzerland}
\author{Dmitri A.~Ivanov}
\affiliation{Institute of Theoretical Physics,
Ecole Polytechnique F\'ed\'erale de Lausanne (EPFL), 
CH-1015 Lausanne, Switzerland}

\date{March 25, 2011}

\begin{abstract}
We study vison excitations in a quantum dimer model
interpolating between the Rokhsar--Kivelson models on 
the square and triangular lattices. In the square-lattice case, the model
is known to be critical and characterized by U(1) topological 
quantum numbers. Introducing diagonal dimers brings the model to 
a Z$_2$ resonating-valence-bond phase. We study variationally
the emergence of vison excitations at low concentration of 
diagonal dimers, close to the critical point. We find that, in this regime, 
vison excitations are large in size and their structure
resembles vortices in type-II superconductors.
\end{abstract}

\maketitle

\section{Introduction}

\revision{The} resonating-valence-bond (RVB) state 
is one of the most exciting proposals
\revision{for} 
strongly correlated phases in two dimensions. 
\revision{Of potential relevance
to} frustrated spin systems \cite{1974FazekasAnderson,2005MisguichLhuillier} 
and high-temperature 
superconductivity \cite{1987Anderson,2004AndersonEtAl},
it \revision{has evaded} a direct identification in experiments and
in studies of realistic models. However, a spinless analogue of the
RVB phase can be systematically studied in quantum dimer models, where
this phase is rigorously confirmed and accessible to a variety of analytic
and numerical methods \cite{2001MoessnerSondhiFradkin,2008MoessnerRaman}.
\revision{Furthermore, quantum dimer models have also motivated
the construction of SU(2) invariant examples of RVB phases in spinful 
sytstems \cite{2005RamanMoessnerSondhi,2005Fujimoto,2010CanoFendley}.
}

The main strongly correlated \revision{feature} 
of the RVB phase \revision{is the existence of}
fractionalized excitations. In a spin system, in the conjectured
RVB phase, two types of elementary excitations must be present: spinons
(spin-1/2 excitations) and visons (Z$_2$ vortices). They have
relative semionic statistics: \revision{a} $-1$ factor for an elementary
braiding of a spinon around a vison \cite{1987KivelsonRokhsarSethna,%
2001SenthilFisher}. In dimer models,
spinon excitations are absent (or effectively pushed to infinitely
high energy)\footnote{\revision{The role of spinon excitations in quantum
dimer models is played by monomers: they can be included by extending
the dimer model. Such extensions are relevant 
for possible connections with the theory of high-temperature 
superconductivity \cite{1987Anderson,2004AndersonEtAl,%
1988RokhsarKivelson}; the deconfinement
of monomers may also serve as a test for the liquid 
phase \cite{2008MoessnerRaman,2002FendleyMoessnerSondhi}.}}, 
and only vison excitations and their combinations
\revision{appear in} the spectrum \cite{2004Ivanov}. 
Another consequence of the existence of
vison excitations is the so called ``topological order'': a degeneracy
of the ground states (and low-lying excitations) for systems on
multiply connected domains \cite{1988RokhsarKivelson,1989ReadChakraborty}.

An interesting question arises \revision{regarding} 
the fate of these elementary
excitations near phase transitions from the RVB phase to neighboring
non-topological phases. Two generic phase transitions of this sort
may be envisioned. Either singlets (or dimers in dimer models)
order into a valence-bond-crystal, or spins order into some
sort of magnetic order. These two \revision{possible paths for} 
the RVB state were
identified in Ref.~\onlinecite{2006Ivanov} as two conditions for
the topological order in Gutzwiller-projected wave functions. 
A field-theoretic  model of \revision{these} phase transitions 
\revision{has} been proposed and analyzed in Ref.~\onlinecite{2009XuSachdev}.

In the present paper, we consider another, non-generic phase
transition into a U(1) critical state. This transition is most
easily realized by deforming a model on a bipartite lattice
(with U(1) winding numbers) into one on a non-bipartite lattice
(with the U(1) symmetry broken down to Z$_2$). While this
transition represents a non-generic, ``fine-tuned'' situation,
it also allows for an analytic treatment of quantum dynamics
near the critical point. Specifically, we consider an interpolation
between the Rokhsar--Kivelson (RK) models on the square and triangular
lattices in such a way that the exact solvability of the RK
ground states is preserved throughout the interpolation.

In such a model, close to the critical point,
\revision{we can variationally construct vison excitations},
thus estimating the size of the vison gap
and identifying the associated length scale (the vison size).
We find that the vison gap closes at the transition point
while the vison size diverges. The shape of the vison in this limit
resembles vortices in type-II superconductors, with the vison size
corresponding to the magnetic penetration length.

The paper is organized \revision{as follows}. In Section II, we briefly
review the properties of the RK dimer models on the square and triangular
lattices.
In Section III, we describe the interpolating model which connects
those lattices. In Section IV, we construct variational
vison excitations near the critical point and discuss their properties.
Finally, in Section V we discuss some interesting
aspects of our results.

\section{Rokhsar--Kivelson dimer models on square and triangular lattices}

\def\horparallel{
\lower.5ex\hbox{
\includegraphics[width=2ex]{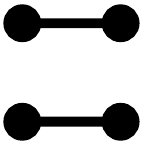}}\,\, }

\def\vertparallel{
\lower.5ex\hbox{
\includegraphics[width=2ex]{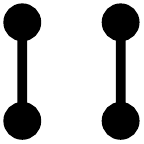}}\,\, }

The Rokhsar--Kivelson dimer model on any lattice is defined in the
Hilbert space of all fully packed coverings of the lattice by dimers
by the Hamiltonian \cite{1988RokhsarKivelson,2001MoessnerSondhi}
\begin{equation}
H_{\rm RK} =\sum \Big( -t \left| \horparallel  \right\rangle\left\langle 
\vertparallel \right|
+ v
\left| \horparallel  \right\rangle\left\langle \horparallel \right|
\Big)
\label{RK-hamiltonian}
\end{equation}
where the sum is taken over all tetragonal plaquettes of the lattice.
In the present work, we are interested in this model on the square and
triangular lattices in two dimensions, in which cases the
sums are understood as those over all the plaquettes of the square lattice
and over all the two-plaquette rhombi of the triangular
lattice, respectively.

On both lattices, there is a special case of the model (\ref{RK-hamiltonian})
with $v=t$ (called the {\it RK point}), in which the ground state is
exactly known and given by all the allowed dimer configurations
with equal amplitudes \cite{1988RokhsarKivelson},
\begin{equation}
\left| {\rm GS} \right\rangle = \frac{1}{\sqrt{Z}}
\sum_c \left| c \right\rangle
\end{equation}
(here the sum is taken over all dimer configurations, and $Z$ is a
normalization factor). However,
the properties of this state and of the lowest excitations above it are
very different in the cases of square and triangular lattices, due
to the fact that one of them (square) is bipartite, and the other one
(triangular) is not.

On the square (bipartite) lattice, the correlations are 
power-law \cite{1963FisherStephenson}
and the spectrum is gapless: the elementary excitations are so called
{\it resonons} with the dispersion 
$\omega(k)\propto k^2$ \cite{1988RokhsarKivelson,1997Henley}.

On the triangular (nonbipartite) lattice, the correlations decay
exponentially with 
distance \cite{2001MoessnerSondhi,2002FendleyMoessnerSondhi,%
2002IoselevichIvanovFeigelman},
and the spectrum is gapped \cite{2001MoessnerSondhi,2004Ivanov} with 
two types of excitations. One sector of excitations is of the {\it vison}
type (topological Z$_2$ vortices), and the other sector is formed by
excitations local in terms of dimer operators (those excitations can
be thought of as composed of an even number of vison excitations).

This difference between the physics of the RK states on bipartite and
non-bipartite lattices can be traced down to a larger set of conserved
quantities in bipartite dimer models. Namely, dimer configurations
on bipartite lattices may be mapped onto a scalar height 
field \cite{1990Levitov,1997Henley,2009NogueiraNussinov},
whose winding numbers provide integer-valued topological invariants
(in the dual variables, those integer winding numbers may be related
to a U(1) gauge symmetry). On the other hand, on non-bipartite lattices,
only a Z$_2$ subgroup of winding numbers survives, thus breaking the
U(1) symmetry down to Z$_2$. The above difference between bipartite
and non-bipartite lattices is well known and extensively discussed
in literature, and we refer the reader to other studies for 
details \cite{2001MoessnerSondhiFradkin,2000SachdevVojta,2008LaeuchliCapponiAssaad}.

\begin{figure}
	\includegraphics[width=0.45\textwidth]{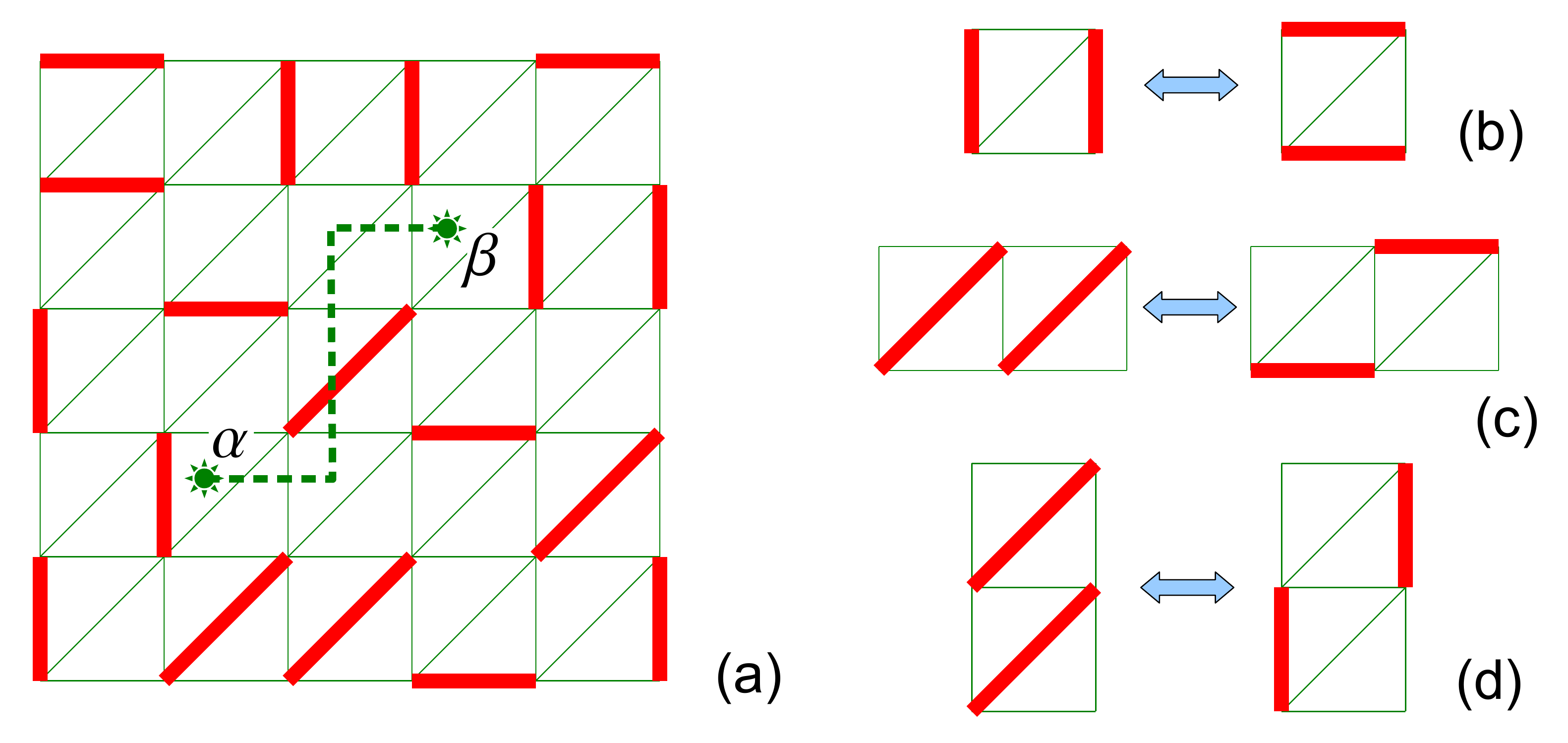}
\caption{{\bf (a)} The interpolated model is defined on the square
lattice with links added along one of the diagonal directions.
The dashed line connecting plaquettes $\alpha$ and $\beta$ illustrates
the \revision{point-vison construction (\ref{two-visons})}.
{\bf (b)--(d)} The three possible dimer-flip processes. The process
(b) is contained in the ``square'' part of the Hamiltonian (\ref{ham-2});
the processes (c) and (d) are included in the ``diagonal'' part
(\ref{ham-3}).}
\label{fig:dimers}
\end{figure}

\section{Interpolation between the RK dimer models on square and triangular
lattices}

\def\diagone{
\lower.45ex\hbox{
\includegraphics[width=4ex]{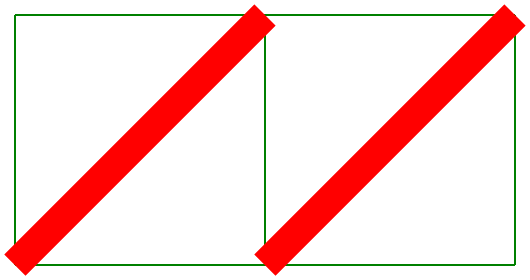}}\,\, }

\def\diagtwo{
\lower.45ex\hbox{
\includegraphics[width=4ex]{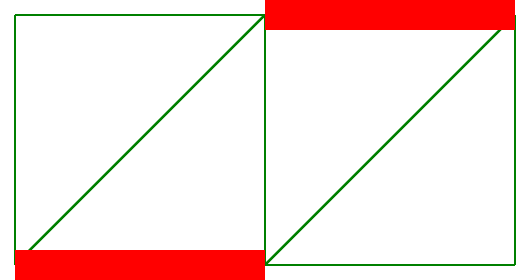}}\,\, }

\def\squareone{
\lower.45ex\hbox{
\includegraphics[width=2ex]{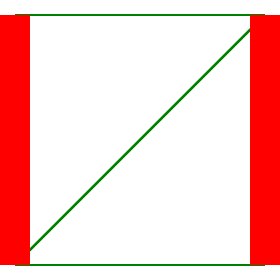}}\,\, }

\def\squaretwo{
\lower.45ex\hbox{
\includegraphics[width=2ex]{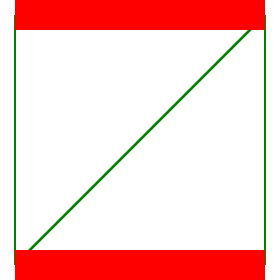}}\,\, }


In our present work, we analyse an interpolation between the RK points
on the square and triangular quantum lattices. We introduce this
interpolation in such a way that the ``RK solvability'' for the ground
state is preserved in the interpolating model. Namely, we construct
the interpolation by introducing dimers along one of the diagonal directions
of the plaquettes of the square lattice (Fig.~\ref{fig:dimers}a). 
Then by tuning the fugacity of
diagonal dimers, one can interpolate between the two limiting cases of the
square and triangular lattices.

The Hamiltonian of the interpolating model is thus chosen to be
\begin{equation}
H=H_{\rm square} + H_{\rm diag}\, ,
\label{ham-1}
\end{equation}
where
\begin{equation}
H_{\rm square}= \sum\Big(
| \squareone \rangle -  | \squaretwo \rangle \Big) 
\Big(  \langle \squareone | - \langle \squaretwo | \Big)
\label{ham-2}
\end{equation}
is the RK Hamiltonian (\ref{RK-hamiltonian}) on the square lattice
with $v=t=1$ (involving kinetic and potential processes for square
plaquettes depicted in Fig.~\ref{fig:dimers}b) and 
\begin{equation}
H_{\rm diag}=\kappa \sum \Big(
| \diagone \rangle - \mu | \diagtwo \rangle \Big) 
\Big(  \langle \diagone | - \mu \langle \diagtwo | \Big)
\label{ham-3}
\end{equation}
is the term involving diagonal dimers. The sum is taken over two
types of parallelogram plaquettes depicted in panels c and d of 
Fig.~\ref{fig:dimers}. The ground state of this Hamiltonian is then
\begin{equation}
\left| {\rm GS} \right\rangle = \frac{1}{\sqrt{Z}}
\sum_c \mu^{D(c)/2} \left| c \right\rangle\, ,
\label{interpolating-GS}
\end{equation}
where $D(c)$ is the  number of diagonals in 
a given configuration $c$.

Note that this is a two-parameter interpolation: the parameter $\mu$
is responsible for the ground state of the Hamiltonian, while the parameter
$\kappa$ determines the dynamics of diagonal dimers. At $\mu=\kappa=0$,
the Hamiltonian reduces to the RK point on the square lattice, while 
$\mu=\kappa=1$ reproduces the RK point on the isotropic triangular
lattice.

The above model was introduced in Ref.~\onlinecite{2006Perseguers}, 
where a simple
variational ansatz for the excitations was proposed. The properties
of the ground state (\ref{interpolating-GS}) were analyzed in
the earlier work Ref.~\onlinecite{2002FendleyMoessnerSondhi}. 
From the Pfaffian technique \cite{1963Kasteleyn}, one easily
deduces the ground-state correlation length $\xi\propto\mu^{-1}$. One
can expect that at any nonvanishing $\mu$ and $\kappa$, the system
belongs to the Z$_2$ phase with excitations continuously connected to those
on the isotropic triangular lattice. This scenario has also been proposed
in Refs.~\onlinecite{2001MoessnerSondhiFradkin} and \onlinecite{2000SachdevVojta}
on the basis of a field-theoretical analysis: such an interpolation
can be interpreted as a U(1) gauge theory coupled to a charge-two scalar field.
In the present paper, we complement this general analysis with a microscopic
construction of vison excitations in the interpolating model
(\ref{ham-1})--(\ref{ham-3}) close to the critical point (for small
deformation parameters $\mu$ and $\kappa$).

In Ref.~\onlinecite{2006Perseguers}, 
variational constructions for dimer (resonon) and vison 
excitations have been proposed. For the resonon excitations, a 
quasi-perturbative
construction gives a gap $\Delta_{\rm dim} \propto \mu^2 \kappa$ (which
is likely to be an exact asymptotic behavior of the dimer gap in our model).
At the same time, an attempt to construct variational vison
excitations as plane waves of {\it point visons} proposed in 
Ref.~\onlinecite{2006Perseguers}
produced states with a finite variational energy that does not tend to
zero in the limit of the square lattice ($\mu,\kappa \to 0$). This
suggests (and we will confirm it in the next section) that point visons
are not good variational states, but the actual vison eigenstates
extends over many lattice spacings (in the limit $\mu,\kappa \to 0$).
The following section is devoted to constructing such states.

\section{Variational construction of vison excitations}
\label{section:variational}

To construct variational visons, we first define {\it point visons} and then
dress them with local operators to approximate eigenstates. To define visons,
one takes a contour $\Gamma_{\alpha\beta}$ connecting two plaquettes 
$\alpha$ and $\beta$ of
the lattice\footnote{Here and below we use greek indices to label plaquettes
of the lattice and latin indices to label lattice sites.} 
and considers the intersection-parity operator 
(Fig.~\ref{fig:dimers}) \cite{1989ReadChakraborty},
\begin{equation}
V_{\alpha\beta}=
(-1)^{{\rm number~of~dimers~intersecting~}\Gamma_{\alpha\beta}} \, .
\label{two-visons}
\end{equation}
One easily verifies (i) that this operator is independent of the contour
$\Gamma_{\alpha\beta}$ connecting two given points, up to an overall controlled
change of sign; (ii) that the commutator of $V_{\alpha\beta}$ 
with any local dimer
Hamiltonian is concentrated near the points $\alpha$ and $\beta$; 
and (iii) that 
$V_{\alpha\beta} V_{\beta\gamma} = V_{\alpha\gamma}$. 
Therefore one can decompose the operator $V_{\alpha\beta}$
into the product of two vison operators, which have the structure
of Z$_2$ vortices \cite{2004Ivanov},
\begin{equation}
V_{\alpha\beta}=V_\alpha V_\beta \, .
\end{equation}
Alternatively, one can understand the operator $V_\alpha$ as the parity
operator (\ref{two-visons}) with the point $\beta$ sent to infinity.

Such a {\it point-vison} operator produces a state orthogonal to the
original state, but not an eigenstate. To form an eigenstate (or,
more precisely, a wave packet of lowest-energy eigenstates), one needs
to {\it dress} a point vison,
\begin{equation}
\left| {\rm V} \right\rangle = 
\hat{D}\, V_\alpha \left| {\rm GS} \right\rangle
\label{variational-vison}
\end{equation}
where $\hat{D}$ is some operator local in terms of dimers. 

We may suggest (and this suggestion is further confirmed by a variational
calculation) that, in order to lower the variational energy, 
we should spread the
vison flux from one plaquette to a certain region of the lattice. On the
square lattice, the point vison may be written in terms of the height
field as
\begin{equation}
V_\alpha = \exp (i K h_\alpha)
\end{equation}
where $K$ is some constant depending on the normalization convention
for the height field $h_\alpha$. The spreading of the vison flux can then be
achieved by the dressing operator
\begin{equation}
\hat{D} = \exp \left[ i \sum_\beta f_\beta (h_\beta- h_\alpha) \right]
\label{dressing-height}
\end{equation}
with some weights $f_\beta$ determining the 
flux redistribution. We will further
use another expression equivalent to Eq.~(\ref{dressing-height}) on the
square lattice, but also directly generalizable to the interpolating model
(where the height field $h_\alpha$ is no longer defined). Namely, we
use the following variational ansatz:
\begin{equation}
\hat{D} = \exp \left[ i \sum_{\langle jk \rangle} 
\tilde{A}_{jk} n_{jk} \right] \, ,
\label{dressing}
\end{equation}
where the sum is taken over all lattice links and $n_{jk}$ is the
dimer density on the link $\langle jk \rangle$ (equal to 1 or 0 in
the presence/absence of a dimer, respectively). Thus constructed variational
state has a gauge redundancy: a transformation
\begin{equation}
\tilde{A}_{jk} \mapsto \tilde{A}_{jk} + \varphi_j + \varphi_k
\label{dressing-gauge}
\end{equation}
does not change the state (\ref{variational-vison}), up to an overall
phase. Therefore, without loss of generality, we may put $\tilde{A}_{jk}=0$
on all diagonal links \revision{and thereby fix the gauge}. 
Finally, it will be convenient to convert
$\tilde{A}_{jk}$ (defined now on the square lattice) 
into a vector potential $A_{jk}$ by
\begin{equation}
A_{jk} = (-1)^j \tilde{A}_{jk}
\end{equation}

The variational energy of the ``square'' part of the Hamiltonian 
(\ref{ham-2}) in the state 
(\ref{variational-vison}), (\ref{dressing}) is given by
\begin{equation}
E_{\rm square} =
\left\langle {\rm V} \right| H_{\rm square} \left| {\rm V} \right\rangle = 
P_{\rm flip}^{(1)} \sum_\beta (1 - \cos \phi_\beta)
\label{energy-square-lattice}
\end{equation}
where $P_{\rm flip}^{(1)}$ is the probability of a given square plaquette to be
flippable (at the RK point on the square lattice,
$P_{\rm flip}^{(1)}=1/4$) \cite{2006Perseguers}
and $\phi_\beta$ is the flux at the plaquette $\beta$.
This flux is composed of the flux of the vector potential $A_{jk}$ and of
the flux of the original point vison $V_\alpha$ in 
Eq.~(\ref{variational-vison}):
\begin{equation}
\phi_\beta=A_{12}+A_{23}+A_{34}+A_{41}+\pi\delta_{\alpha\beta}\, ,
\end{equation}
where the numerical indices 1 to 4 label the four sites of the square plaquette
$\beta$ in the cyclic order.
Since we assume that the operator $\hat{D}$ is local in terms of dimers and
concentrated in a finite region near the plaquette $\alpha$, the total flux
of the vector potential $A_{jk}$ must vanish. Therefore, we find the constraint
for the total flux $\phi_\beta$,
\begin{equation}
\sum_\beta \phi_\beta =\pi\, .
\label{total-flux}
\end{equation}

On the square lattice (at $\mu=\kappa=0$), the total variational energy
is given by Eq.~(\ref{energy-square-lattice}). Its minimization with the
constraint (\ref{total-flux}) yields the spreading of the flux over the whole
lattice. Indeed, spreading the flux over $N$ plaquettes gives $N$ fluxes
$\phi_\beta \propto 1/N$, which results in the total energy $E \propto 1/N$.
This variational construction shows that, indeed, in the limit of the square
lattice the vison excitation becomes unstable and that 
close to the critical point
the vison size must be large (much larger than the lattice spacing). In this
limit, using the fact that the optimal fluxes $\phi_\beta$ are small, we
can expand the cosine in Eq.~(\ref{energy-square-lattice}) and write
the long-wavelength expression for the energy,
\begin{equation}
E_{\rm square} = \frac{P_{\rm flip}^{(1)}}{2} \int 
(\vec{\nabla} \times \vec{A})^2\, d^2 x \, .
\label{energy-square}
\end{equation}
with the total screened flux
\begin{equation}
\oint_{x_\alpha} \vec{A}\, d\vec{x} = \pi
\end{equation}
[the latter integral is taken over a small circle surrounding
the position $x_\alpha$ of the point vison in Eq.~(\ref{variational-vison});
this point is excluded from the integral (\ref{energy-square})].

In the interpolating model (at finite $\mu$ and $\kappa$), the second
term in the Hamiltonian (\ref{ham-3}) stabilizes the vison. Indeed, its
contribution for our variational state (\ref{variational-vison}), 
(\ref{dressing}) can be written, in the long-wavelength limit, as
\begin{equation}
E_{\rm diag} =4 \kappa\mu^2 P_{\rm flip}^{(2)} \int \vec{A}^2\, d^2 x \, .
\label{energy-diag}
\end{equation}
Here we assume that the vector potential $\vec{A}$ is small and slowly
varying on the length scale of one lattice constant. This is a valid
assumption far from the vison center. The central region gives a contribution
of the order $\kappa\mu^2$ to the vison energy, which can be neglected
to the leading order, in comparison with the logarithmically larger
contribution from large distances. The coefficient
$P_{\rm flip}^{(2)}$ is given by the probability to find two parallel
non-diagonal dimers on a ``skew'' (parallelogram) plaquette. To the leading
order in $\mu$, it can be approximated by its value at the RK point
on the square lattice, 
$P_{\rm flip}^{(2)}\approx 0.04542$ \cite{2006Perseguers}.
The total variational energy of the vison can now be written as
\begin{equation}
E=E_{\rm square} + E_{\rm diag}
=\frac{P^{(1)}_{\rm flip}}{2} \int d^2 x \, 
\left[ (\vec{\nabla} \times \vec{A})^2 + \frac{1}{L^2} \vec{A}^2 \right]\, ,
\label{total-energy-A}
\end{equation}
where the new large length scale (``size of vison'') emerges,
\begin{equation}
L=\sqrt{ \frac{P^{(1)}_{\rm flip}}{8 P^{(2)}_{\rm flip}}}
\cdot \kappa^{-1/2} \mu^{-1}\, .
\end{equation}

Remarkably, this variational problem is equivalent to that of a vortex
in a type-II superconductor (with $L$ corresponding to the London penetration
length)\footnote{\revision{This analogy assumes a vortex
structure translationally invariant along the vortex core (i.e., either
at zero temperature or with the infinite stiffness in the third dimension)
and does not imply any similarities in thermodynamic properties. At
a more qualitative level,} an anlogy between visons and 
superconducting vortices has also been noted previously in the context of the
connection between the RVB phase and 
superconductivity \cite{ReadSachdev,2002IvanovSenthil}.} \cite{1996Tinkham}.
 In terms of the ``magnetic field'' defined as
\begin{equation}
B=\vec{\nabla}\times \vec{A}\, ,
\end{equation}
the variational problem (\ref{total-energy-A}) is equivalent to
minimizing the energy
\begin{equation}
E=\frac{P^{(1)}_{\rm flip}}{2} \int d^2 x \, 
\left[ B^2 + L^2 \left( \vec{\nabla}\times B \right)^2 \right]\, ,
\label{total-energy-B}
\end{equation}
with the total-flux constraint
\begin{equation}
\int B\, d^2 x = \pi \, .
\end{equation}

One can find a centrally-symmetric solution to this variational
problem,
\begin{equation}
B(R) = \frac{1}{2 L^2} K_0 \left(\frac{R}{L}\right)\, ,
\label{vortex-shape}
\end{equation}
where $K_0$ is the modified Bessel (Macdonald) function
and $R$ is the distance from the vison center. The
energy corresponding to this solution is
\begin{equation}
E= \frac{\pi P^{(1)}_{\rm flip}}{4 L^2} \ln L = 
-\pi P^{(2)}_{\rm flip} \kappa \mu^2 \ln (\kappa \mu^2)\, .
\end{equation}
Of course, this result for the energy, as well as the long-wavelength
approximation (\ref{energy-square}) -- (\ref{total-energy-A}) and 
(\ref{total-energy-B}) -- (\ref{vortex-shape}), is only justified if $L \gg 1$.
The expression (\ref{vortex-shape}) for the variational
solution is also only valid for $R \gg 1$.

\section{Discussion of results}

In the preceding section, we have constructed a variational vison excitation
in a near-critical RK-type dimer model interpolating 
between the RK points on the square and triangular lattices. While, 
rigorously speaking, our construction provides an upper bound for the gap
in the vison excitation sector, we believe that it captures the main
qualitative feature of vison excitations (spreading the vison flux
over a large region of lattice), and therefore conjecture that it also
gives the exact functional dependence on the deformation parameters $\mu$
and $\kappa$. Below we discuss several interesting aspects and implications
of our result.

First of all, our variational construction is only based on the
large length scale $L \gg 1$, which is equivalent to $\kappa\mu^2 \ll 1$.
Thus we do not, in fact, require the individual smallness of the
parameters $\kappa$ and $\mu$, but only of their above combination.

Second, we find that the (variational) vison gap closes as 
\begin{equation}
E_{\rm vison} \propto \frac{1}{L^2} \ln L
\label{energy-short-vison}
\end{equation}
at the critical point. This behavior should be compared to the
variational analysis of the dimer gap in the same model performed
in Ref.~\onlinecite{2006Perseguers}. There it was found that the 
dimer gap closes as
\begin{equation}
E_{\rm dimer} \propto \frac{1}{L^2}\, ,
\label{energy-short-dimer}
\end{equation}
i.e., it is always smaller close to the critical point. This suggests
that visons tend to
form bound pairs thus canceling the logarithmic term in the energy
(see also the discussion of two vison signs below).

Third, our variational vison can have two different signs, which correspond
to the two opposite signs of the smeared flux $B$. While there is only
one species of the point vison (fluxes $\pi$  and $-\pi$ on one plaquette 
are identical), there is a freedom of smearing this flux as either positive
or negative over many plaquettes (of course, there are also options to
smear fluxes $\pm 3\pi$, $\pm 5\pi$, etc., but they are obviously higher in
energy). As a result, we can construct two variational visons differing
by the sign of the field $B$. This result may appear surprising in view
of our expectation of the Z$_2$ structure of vison excitations. This
paradox can be resolved by analysing the properties of our variational
visons with respect to the translational symmetries.

\begin{figure}
\includegraphics[width=0.45\textwidth]{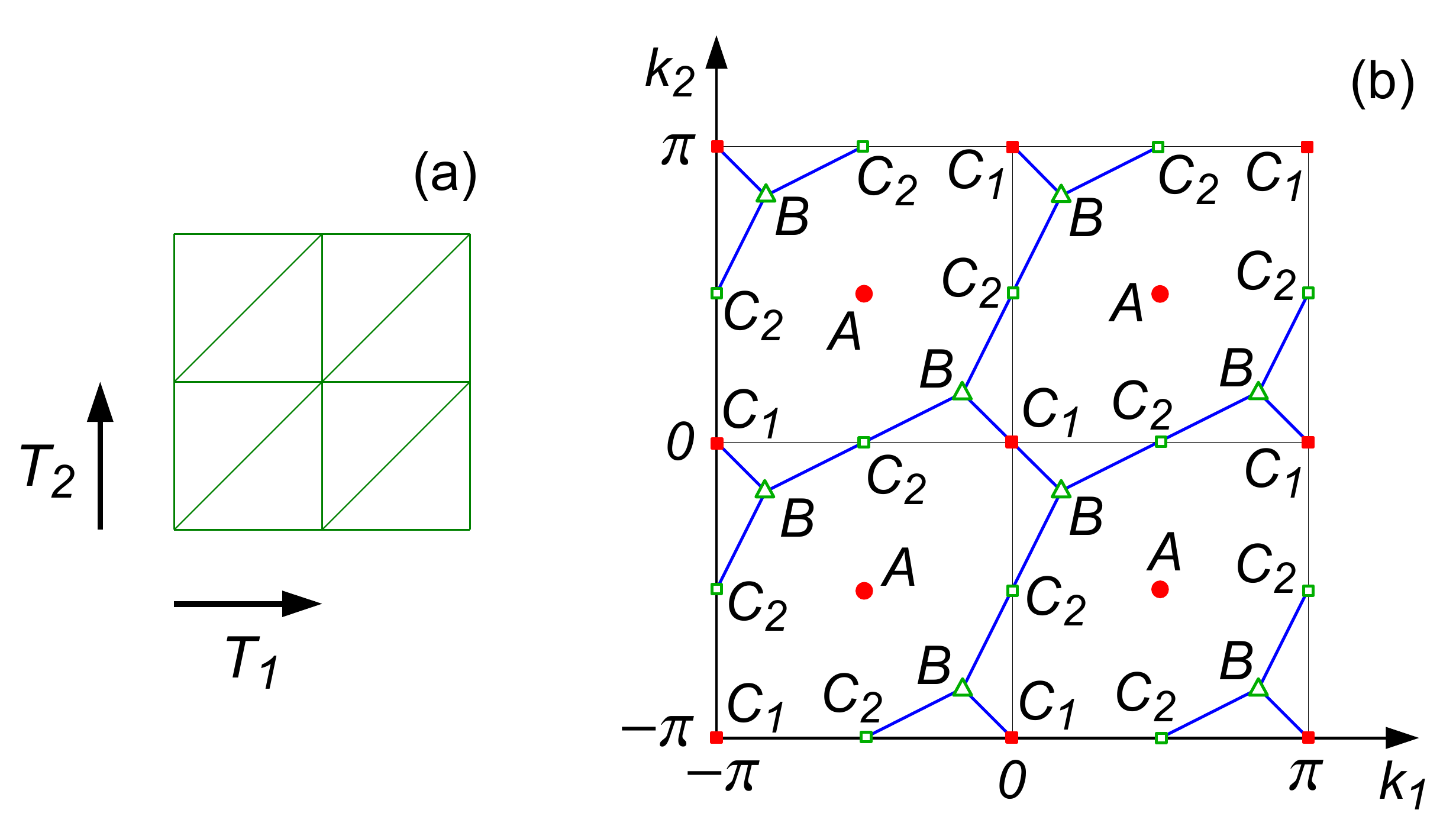}
\caption{{\bf (a)} The two elementary translations of the lattice,
which anticommute for vison states. {\bf (b)} The $k$ space for
vison excitations. The labeling of the points $A$, $B$, and $C$ is
the same as in the isotropic triangular lattice case considered
in Ref.~\onlinecite{2004Ivanov} (see Fig.~3 there). 
In our interpolating model, 
points $C_1$ and $C_2$ are not degenerate: they only become degenerate 
in the triangular lattice limit ($\kappa=\mu=1$). Close to the critical
point (square lattice), the vison minima are located at points $C_1$.
Note that the vison Brillouin zone is only one half of the usual
dimer Brillouin zone, and therefore it contains two $C_1$ points.}
\label{fig:visons}
\end{figure}

As follows from their construction, visons live on a frustrated 
lattice \cite{2001MoessnerSondhiFradkin,2004Ivanov}. In terms
of translational symmetries, this implies that the elementary translations
in the $x_1$ and $x_2$ directions (along the links of the square lattice,
Fig.~\ref{fig:visons}a)
anticommute:
\begin{equation}
T_1 T_2 = - T_2 T_1
\end{equation}
Therefore, to define a wave vector for vison eigenstates, one needs to
double the unit cell. In terms of $k$ vectors, this implies the period
of $\pi$ in the directions $k_1$ and $k_2$ for the dispersion relation.
More precisely, to the {\it four} $k$ points $(k_1,k_2)$, $(k_1+\pi,k_2)$,
$(k_1,k_2+\pi)$, and $(k_1+\pi,k_2+\pi)$, there correspond {\it two}
linearly independent vison eigenstates degenerate in energy.

By analyzing the action of the translation operators on our variational
vison states, we conclude that these states belong to the vicinity of the
wave vectors generated by $k=(0,0)$. Thus the two vison
states of opposite fluxes correspond to the two degenerate minima of
the vison dispersion relation. Extending the notation of 
Ref.~\onlinecite{2004Ivanov}, these
minima can be denoted as $C_1$ (which become degenerate with points
$C_2$ in the isotropic triangular limit, see Fig.~\ref{fig:visons}b). 
Note that these points
are different from the energy minima on the isotropic triangular lattice
($\kappa=\mu=1$), which are located at points $B$ \cite{2004Ivanov}.

Remarkably, the existence of visons of opposite signs can qualitatively
explain the formation of vison bound states \revision{(dimer-like
excitations) found numerically in 
Ref.~\onlinecite{2004Ivanov}}. Indeed, superimposing
visons of opposite signs at a short distance (much shorter than $L$)
compensates the long-distance logarithmic contribution to the energy
(\ref{energy-short-vison}), thus producing the energy of the bound state
(\ref{energy-short-dimer}).

Fourth, an analogy with type-II superconductors can be drawn not only in the
variational form of the energy (\ref{total-energy-A}) or
(\ref{total-energy-B}), but also in the existence of two different length
scales. The vison size $L$ in our problem obviously corresponds to the
London penetration length in the theory of superconductivity. On the
other hand, there is a second length scale $\xi$ (the ground-state
correlation length), which resembles the superconducting coherence
length. In superconductors, the interplay of the two length scales
leads to important physical consequences (difference between type-I
and type-II superconductors). We may therefore conjecture that
in quantum dimer models some novel effects depending on the relation
between the two lengths may also be possible. Note, however, that
this analogy is not complete: in our vison construction, 
the length scale $\xi$ does not enter, and only 
the length $L$ affects the variational state
and its energy. In fact, the lower cut-off of the logarithm 
in the vison energy (\ref{energy-short-vison}) is given by the
lattice spacing rather then $\xi$. It would be interesting
to explore further improvements of our variational ansatz, 
which may bring in the length scale $\xi$ (by analogy
to the core of superconducting vortices). We leave this 
interesting question for future study.

Another interesting question that remains unexplored is the dynamics
of visons in our model. While exact vison eigenstates must carry a
well-defined wave vector $k$, our variational state is localized in
space and thus corresponds to a wave packet of width $\Delta k \sim 1/L \ll 1$
in the reciprocal space. We believe that its energy accurately
represents the bottom of the vison band, but the question of the
vison mass remains unresolved in our construction.

While the study reported in the present paper relates to a very specific
model and to a non-generic phase transition between a RVB phase
and a U(1) critical point, we believe that some of our findings may
be generalized to other systems and situations. In particular,
the existence of two different length scales ($\xi$ and $L$) is likely
to be a general property of the RVB phase. The divergence of the
vison size may also occur at other types of second-order
phase transitions. We therefore hope that the example considered
in the present paper will provide a guidance for future studies
of the RVB phase in various systems and models.

\medskip

We are grateful to G.~Blatter, M.~Feigelman, R.~Moessner, 
D.~Poilblanc, and D.~Sadri for helpful discussions and comments on this work.

\end{document}